\documentclass[secnumarabic,aps,pra,amsfonts,
twocolumn,
showkeys,
floatfix%
]{revtex4}

\usepackage{bm,amsfonts}
\usepackage{graphicx}

\newcommand{\gl}[1]{(\ref{#1})}

\begin{document}
\title{Superconducting band stabilizing superconductivity in MgB$_2$} 

\author{Ekkehard Kr\"uger}


\affiliation{Institut f\"ur Metallkunde, 2. Lehrstuhl, Universit\"at Stuttgart,
 D-70569 Stuttgart, Germany}

\begin{abstract}
  It is shown that the superconducting intermetallic compound MgB$_2$ possesses
  a narrow, partly filled ``superconducting band'' with Wannier functions of
  special symmetry in its band structure. This result corroborates previous
  observations about the band structures of numerous superconductors and
  non-superconductors showing that evidently superconductivity is always
  connected with such superconducting bands. These findings are interpreted in
  the framework of a nonadiabatic extension of the Heisenberg model. Within
  this new group-theoretical model of correlated systems, Cooper pairs are
  stabilized by a nonadiabatic mechanism of constraining forces effective in
  narrow superconducting bands. The formation of Cooper pairs in a
  superconducting band is mediated by the energetically lowest boson
  excitations in the considered material that carry the crystal spin-angular
  momentum $1\cdot\hbar$. These crystal-spin-1 bosons are proposed to
  determine whether the material is a conventional low-$T_c$ or a high-$T_c$
  superconductor. This interpretation provides the electron-phonon mechanism
  that enters the BCS theory in conventional superconductors.
\end{abstract}
\maketitle
\section{Introduction}
\label{sec:intro}
In a former paper~\cite{enhm} the author proposed to generalize the classical
Heisenberg model of magnetism~\cite{hei} by introducing three new postulates
emphasizing the \emph{atomic-like} motion of electrons in narrow, partly filled
energy bands. The resulting ``nonadiabatic Heisenberg model'' (NHM) uses the
term ``atomic-like motion'' in the sense of Mott \cite{mott} and
Hubbard\cite{hubbard}: as long as possible the electrons occupy localized
states and perform their band motion by hopping from one atom to another.
 
This group-theoretical model provides a novel concept to understand
correlation effects in narrow bands in terms of symmetry-adapted Wannier
functions and was developed to interpret the existence of magnetic and
superconducting bands with Wannier functions of special symmetries in the band
structures of magnetic and superconducting materials.

Evidently, the magnetic states in Cr~\cite{ea}, Fe~\cite{ef},
La$_2$CuO$_4$~\cite{josla2cuo4}, and YBa$_2$Cu$_3$O$_6$~\cite{ybacuo6} are
connected with narrow, partly filled magnetic bands in the band structures of
the respective materials. Superconducting bands (as shall be defined in
Sec.~\ref{sec:defsbands}) have already been identified in the band structures
of numerous elemental superconductors~\cite{es2} and of the high-temperature
superconductors La$_2$CuO$_4$~\cite{josla2cuo4} and
YBa$_2$Cu$_3$O$_7$~\cite{ybacuo7}. Furthermore, partly filled superconducting
bands cannot be found in those elemental metals (such as Li, Na, K, Rb, Cs, Ca,
Cu, Ag, and Au) which do not become superconducting~\cite{es2}. An
investigation into the band structures of the transition metals in terms of
superconducting bands straightforwardly leads to the Matthias
rule~\cite{josm}.

The aim of the present paper is to show that the intermetallic compound
MgB$_2$ with the relatively high transition temperature $T_c \approx 39
K$~\cite{nagamatsu} possesses a narrow, partly filled superconducting band in
its band structure.  Before in Sec.~\ref{sec:sband2} the existence of this
band shall be established, the following Sec.~\ref{sec:nhm} outlines the
physical substance of the new nonadiabatic model and Sec.~\ref{sec:sbands}
gives the definition of superconducting bands and a short characterization of
the mechanism of Cooper pair formation in these bands.

\setcounter{equation}{0}
\section{Nonadiabatic Heisenberg model}
\label{sec:nhm}

In this section the nonadiabatic Heisenberg model is shortly
characterized, a detailed substantiation of all the statements is given in
Ref.~\cite{enhm}.

Within the NHM, the Hamiltonian
\begin{equation}
  \label{eq:9}
H^{n} = H_{HF} + H_{Cb}^{n}  
\end{equation}
in a partly filled superconducting or magnetic band consists of the familiar
Hartree-Fock operator $H_{HF}$ and the nonadiabatic Coulomb interaction
\begin{eqnarray}
  \label{eq:10}
H_{Cb}^{n} &=& \sum_{{\bf T}, m}\langle{\bf T}_{1}, m_{1}, n; 
{\bf T}_{2}, m_{2}, n|H_{Cb}|
{\bf T}_{1}', m_{1}', n; {\bf T}_{2}', m_{2}', n \rangle\nonumber\\
&&\times c_{{\bf T}_{1}m_{1}}^{n\dagger}
c_{{\bf T}_{2}m_{2}}^{n\dagger}
c_{{\bf T}_{2}'m_{2}'}^{n}
c_{{\bf T}_{1}'m_{1}'}^{n}.
\end{eqnarray}
The fermion operators $c_{{\bf T}m}^{n\dagger}$ and $c_{{\bf T}m}^{n}$ create
and annihilate electrons in ``nonadiabatic localized states'' $|{\bf T}, m, n
\rangle$ [with crystal spin $m$ (see Sec.~\ref{sec:crystalspin}) on the atom
at position ${\bf T}$] that depend on an additional quantum number $n$ labeling
different states of the nonadiabatic motion of the center of mass of the
localized states~\cite{bem2}. The matrix elements of $H_{Cb}^{n}$ are integrals
over nonadiabatic localized functions as given in Eq.~(2.17) of
Ref.~\cite{enhm}.

The introduction of these nonadiabatic localized states allows a realistic
description of an atomic-like motion of the electrons. While in the framework
of the adiabatic approximation the electrons move in {\em rigid} orbitals in the
{\em average} potential of the other electrons, in the nonadiabatic system a
localized electron moves in a potential depending on which of the adjacent
localized states is occupied and on the present motion of the electrons in
these states. Such a correlated electronic motion within a fluctuating
potential leads to localized orbitals which are not symmetric at any moment,
but only on the average of time. As a consequence, the centers of mass of the
localized states become permanently accelerated in varying directions. The
resulting nonadiabatic motions of the centers of mass are labeled by the
new quantum number $n$.

The NHM assumes that there exist nonadiabatic localized states satisfying the
equation
\begin{equation}
  \label{eq:11}
\langle{\bf T}_{1}, m_{1}, n; 
{\bf T}_{2}, m_{2}, n|H_{Cb}|
{\bf T}_{1}', m_{1}', n; {\bf T}_{2}', m_{2}', n \rangle
= 0
\end{equation}
for 
\begin{equation}
  \label{eq:12}
\{{\bf T}_1,{\bf T}_2\} \neq \{{\bf T}_1',{\bf T}_2'\}    
\end{equation}
if the considered partly filled band is one of the narrowest bands of the
given material, where $\{{\bf T}_1,{\bf T}_2\} = \{{\bf T}_1',{\bf T}_2'\}$ means
${\bf T}_1 = {\bf T}_1'$ and ${\bf T}_2 = {\bf T}_2'$ or ${\bf T}_1 = {\bf T}_2'$
and ${\bf T}_2 = {\bf T}_1'$.

This Eq.~\gl{eq:11} follows from the first postulate of the NHM which assumes
that the correlation effects specified by Eq.~\gl{eq:12} are energetically
unfavorable in narrow, partly filled bands. So the electrons modify their
localized orbitals within the nonadiabatic localized states in such a way that
the transitions specified by Eq.~\gl{eq:12} do not occur. These modified
electronic orbitals yield a well-defined motion of the centers of mass labeled
by the quantum number $n$.

Eq.~\gl{eq:11} {\em defines} the ``atomic-like state of motion'' within the
NHM and replaces the stronger Heisenberg condition claiming that in narrow,
half-filled bands there is exactly one electron on each atom.

\subsection{Novel features of the nonadiabatic Heisenberg model}

A correlated electron system characterized by Eq.~\gl{eq:11} possesses three
novel features distinguishing it from any system described within the
adiabatic approximation. First, within the NHM there exists a ``nonadiabatic
condensation energy'' $\Delta E$; second, the nonadiabatic Hamiltonian $H^n$
possesses unusual commutation properties; and thirdly, the ``crystal spin'' of
the nonadiabatic localized states is a conserved quantity.

\subsubsection{Nonadiabatic condensation energy}

Eq.~\gl{eq:11} is assumed to be satisfied in the {\em ground state} of the
narrowest, partly filled bands of the metals. Since this equation is clearly
not true within the adiabatic approximation, the Coulomb correlation energy of
the nonadiabatic atomic-like state characterized by Eq.~\gl{eq:11} is lower
than the correlation energy of this state within the adiabatic
approximation. Hence, the electrons of the narrowest, partly filled bands of
the metals lower their energy by the ``nonadiabatic condensation energy''
\begin{equation}
  \label{eq:5}
\Delta E = E_{b} - E_{a} 
\end{equation} 
at the transition from the adiabatic (more bandlike) to the nonadiabatic
atomic-like state. $E_a$ and $E_b$ denote the ground-state energies of the
operator $H^n$ in Eq.~\gl{eq:9} and of the related operator within the
adiabatic approximation, respectively.  $\Delta E$ may be approximated by
Eq.~(2.20) of Ref.~\cite{enhm}.

\subsubsection{Commutation properties of the nonadiabatic Hamiltonian $H^n$}

In superconducting and magnetic bands the symmetry of the nonadiabatic
localized states is not adapted to the space group $G$ of the considered
material, but only to a subgroup $M$ of $G$. As a consequence of
Eq.~\gl{eq:11}, the nonadiabatic Hamiltonian $H^n$ in Eq.~\gl{eq:9} commutes
with all the symmetry operators of $M$, {\em but not} with the symmetry
operators of $G$ that do not belong to $M$. Thus, the group-theoretical NHM
allows a straightforward physical interpretation of the symmetry of the
localized states related to the atomic-like motion in narrow, partly filled
bands.

This feature distinguishes $H^n$ from any Hamiltonian $H$ written in
the adiabatic approximation since the symmetry properties of $H$ are
independent of the symmetry of the Wannier basis used to calculate its matrix
elements. 

\subsubsection{Crystal-spin angular momentum}
\label{sec:crystalspin}
The nonadiabatic localized states are no longer labeled by the spin quantum
number $s = \pm\frac{1}{2}$, but by a new quantum number $m = \pm\frac{1}{2}$
which may be called the quantum number of the ``crystal spin''. This is in
analogy to the wave vector ${\bf k}$ of the Bloch functions which is sometimes
referred to as ``crystal momentum'' in order to distinguish it from a true
momentum.

Within the nonadiabatic correlated system the conservation law of spin angular
momentum as expressed by the equation
\begin{equation}
  \label{eq:17}
[H, S(\alpha )] = 0 \quad\mbox{for }\alpha \in O(3)
\end{equation}
is replaced by the conservation law
\begin{equation}
  \label{eq:18}
[H^n, M(\alpha )] = 0\quad\mbox{for }\alpha \in G_M   
\end{equation}
of the crystal spin $m$. The operators $S(\alpha)$ are the symmetry operators
of the electron spin and $O(3)$ stands for the three-dimensional rotation
group; $M(\alpha)$ and $G_M$ denote the analogous operators and the
corresponding group, respectively, in the space group of the considered
material.

At interactions of the electrons with phonons, a Bloch state bears the
crystal-spin angular momentum $m = \frac{1}{2}\cdot\hbar$ and suitable linear
combinations of the phonons carry the crystal spin $m = 1\cdot\hbar$.

\subsection{Superconducting and magnetic bands}
\label{sec:smbands}
Localized functions $\langle {\bf r},t,{\bf q}\,|{\bf T}, m, n\rangle$
representing the nonadiabatic localized states are highly complicated since
they depend on an additional coordinate ${\bf q}$ related to the nonadiabatic
motion of the center of mass (while ${\bf r}$ and $t$ denote, as usual, the
local and spin coordinate, respectively, of the localized
electron). Fortunately, these functions need not to be known. The NHM only
postulates their existence and assumes that they have the same symmetry and
spin dependence as the best localized exact Wannier functions of the considered
partly filled energy band. In this context, ``exact Wannier functions'' form a
{\em complete} basis of the Bloch functions of this band. 

In several narrow, partly filled energy bands of the metals the electrons can
gain the nonadiabatic condensation energy $\Delta E$ only under specific
conditions: in a narrow, partly filled ``magnetic band''~\cite{enhm,ybacuo6}
related to a magnetic structure $S$ the electrons can occupy the atomic-like
state \emph{only if} this magnetic structure $S$ actually exists in the
considered material. Further, electrons in a narrow, partly filled
``superconducting band'' (as defined in Sec.~\ref{sec:defsbands}) occupying the
atomic-like state are forced in a new way to form Cooper pairs below a certain
transition temperature. This ``new way'' will be substantiated in
Sec.~\ref{sec:cooper}.

 This result suggests that the nonadiabatic condensation energy $\Delta E$
 in Eq.~\gl{eq:5} stabilizes both magnetism~\cite{ef,ea} and
 superconductivity~\cite{es,josn}. Both phenomena have the same physical origin:
 they exist because the electrons at the Fermi level tend to occupy the
 energetically favorable atomic-like state. The special symmetry and spin
 dependence of the related Wannier function determine whether the material
 becomes magnetic or superconducting (or has a property not yet considered).
 This important statement of the NHM is corroborated by the calculated band
 structures mentioned in Sec.~\ref{sec:intro}.

 \begin{figure*}[t]
  \includegraphics[width=.55\textwidth,angle=-90]{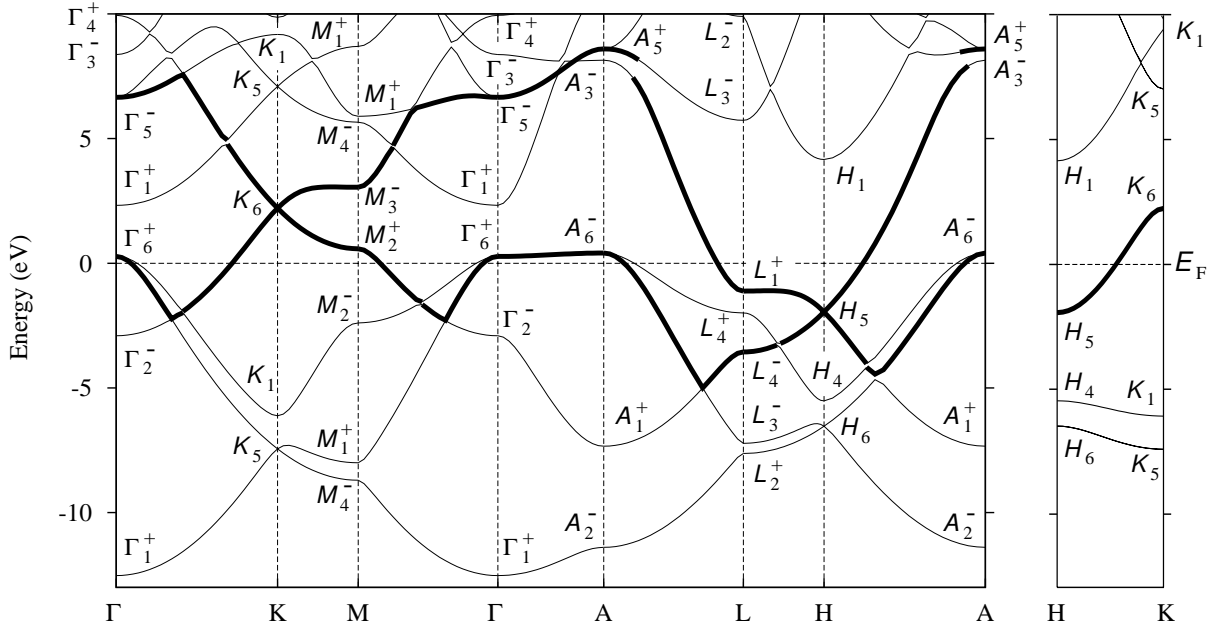}
 \caption{
   Band structure of MgB$_2$ as calculated by Ove Jepsen~\cite{jepsen} with 
   symmetry labels as given in Table~\ref{tab:darst2}. The bold line
   shows the superconducting band. It is related to the B atoms and consists
   of two branches because there are two B atoms in the unit cell.
  \label{fig:bandstr2}
}
 \end{figure*}


\setcounter{equation}{0}
\section{Superconducting bands and superconductivity}
\label{sec:sbands}
\subsection{Definition of superconducting bands}
\label{sec:defsbands}
Usually, the energy bands crossing the Fermi level in the paramagnetic metals
are degenerate at several points and lines of symmetry of the Brillouin
zone. Therefore, it is not possible to separate narrow isolated sets of bands
whose Bloch functions can be unitarily transformed into best localized Wannier
functions that are symmetry-adapted to the full space group $G$ of the
considered metal. However, in some cases such Wannier functions may be
constructed if we allow that they are adapted only to the symmetry of a
magnetic subgroup $M$ of $G$ or if they are allowed to be spin
dependent~\cite{enhm}. In the first case, the band is a magnetic band as
mentioned in the foregoing Sec.~\ref{sec:smbands}, in the second case it is a
superconducting band.

We define an energy band of a given material to be a ``superconducting band''
if the Bloch functions of this band can be unitarily transformed into
\emph{spin-dependent} Wannier functions $w_{im}({\bf r} - {\bf R} - \bm\rho_i,
t)$ (as defined by Eq.~(A22) of Ref.~\cite{enhm}) which are
\begin{itemize}
\item centered on a (well-defined) part of the atoms (at positions ${\bf T} =
  {\bf R} + \bm\rho_i$);
\item symmetry-adapted to the (full) space group $G$ of this material;
\item labeled by the quantum number $m = \pm\frac{1}{2}$ of the crystal spin
  [see note (ii) of Table~\ref{tab:atomwf2}];
  and  
\item localized as well as possible.
\end{itemize}

\subsection{Mechanism of Cooper pair formation in a superconducting band}
\label{sec:cooper}

In a narrow, roughly half-filled superconducting band, the electron system has
no other possibility to gain the nonadiabatic condensation energy $\Delta E$
in Eq.~\gl{eq:5} but to occupy an atomic-like state represented by
\emph{spin-dependent} Wannier functions. This spin dependence has far reaching
consequences because only special atomic-like motions with spin-dependent
localized states may exist in the nonadiabatic system. These states are
determined by the interplay of two conservation laws: on the one hand, the
bare electrons satisfy the conservation of the electron spin and, on the
other hand, the nonadiabatic localized states conserve the crystal spin.

In this section, the atomic-like motion with spin-dependent localized states
is shortly characterized, a detailed substantiation of all the statements is
given in Ref.~\cite{josn}.

\subsubsection{Nonadiabatic atomic-like motion with spin-dependent localized states}
\label{sec:general}
When the localized states are represented by spin-dependent Wannier functions,
the nonadiabatic operator $H_{Cb}^{n}$ of Coulomb interaction complying with
Eq.~\gl{eq:11} does not conserve the crystal spin angular momentum,
i.e., 
\begin{equation}
  \label{eq:28}
[H_{Cb}^n, M(\alpha )] \neq 0
\end{equation}
for at least one $\alpha \in G_M$. At first sight, this result seems to show
that the NHM is not applicable to superconducting bands. However, the
nonadiabatic motion of the centers of mass in the localized states gives point
to another interpretation. Remember that the quantum number $n$ in
Eq.~\gl{eq:11} labels the special nonadiabatic motion of the centers of mass
of those localized states which satisfy this Eq.~\gl{eq:11}. Thus,
Eq.~\gl{eq:28} indicates that this special nonadiabatic motion occurs in such
a way that phonons or other boson excitation are excited (or absorbed) which
store the surplus crystal spin angular momenta.  This interpretation is
corroborated by the fact that by the mere addition of symmetrized boson
operators we may construct from $H_{Cb}^{n}$ an interaction
\begin{eqnarray}
H_{Cb}^{ns} &=& \displaystyle\sum_{{\bf T}, m}
\langle{\bf T}'_{1}, l_{1}; 
{\bf T}'_{2},l_{2};{\bf T}_{1},m_{1},n; 
{\bf T}_{2},m_{2},n|H_{Cb}\nonumber\\
&&
|{\bf T}_{1},m_{1}',n;{\bf
T}_{2},m_{2}',n\rangle
b_{{\bf T}'_{1}l_{1}}^{\dagger}
b_{{\bf T}'_{2}l_{2}}^{\dagger}\nonumber\\
&&
\times
c_{{\bf T}_{1}m_{1}}^{n\dagger}
c_{{\bf T}_{2}m_{2}}^{n\dagger}
c_{{\bf T}_{2}m_{2}'}^{n}
c_{{\bf T}_{1}m_{1}'}^{n} + \mbox{H.c.}
\label{eq:13}
\end{eqnarray}
which conserves the crystal spin,
\begin{equation}
  \label{eq:19}
[H_{Cb}^{ns}, M(\alpha )] = 0\quad\mbox{for }\alpha \in G_M.
\end{equation}
The boson operator $b_{{\bf T}l}^{\dagger}$ creates a localized boson $|{\bf T},
l\rangle$ with crystal spin $l = -1, 0, +1$ at the position ${\bf T}$.

This ``spin-boson interaction'' $H_{Cb}^{ns}$ replaces the Coulomb interaction
$H^n_{Cb}$ in a narrow superconducting band. In cubic crystals the matrix
elements of $H_{Cb}^{ns}$ are determined by Eq.~(4.28) of
Ref.~\cite{es}. The complete nonadiabatic Hamiltonian $H^n$ now may be
written as
\begin{equation}
  \label{eq:14}
H^n = H_{HF} + H_{Cb}^{ns} + H_{b}  
\end{equation}
with $H_{b}$ denoting the operator of the boson energy. 

\subsubsection{Atomic-like motion with spin-dependent localized states at zero
temperature}
\label{sec:zerot}
Since $H_{Cb}^{ns}$ depends on boson operators, a certain number of
crystal-spin-1 bosons is excited in the ground state of the nonadiabatic
Hamiltonian $H^n$ at any temperature. However, at zero temperature we may
assume that these bosons are virtually excited, i.e., each boson pair is
reabsorbed immediately after its generation, producing in this way an effective
{\em electron-electron} interaction. Thus, at zero temperature we approximate
the nonadiabatic system represented by $H^n$ by a purely electronic system
represented by a Hamiltonian
\begin{equation}
  \label{eq:15}
H^0 = H_{HF} + H_{Cb}^0  
\end{equation}
not depending on boson operators.

Also at zero temperature, the nonadiabatic mechanism specified by
Eq.~\gl{eq:11} occurs in the nonadiabatic system. In the purely electronic
system represented by $H^0$, however, the electronic motion is no longer
coupled to the motion of the centers of mass of the localized states, but the
interaction term $H_{Cb}^0$ of $H^0$ contains the effective electron-electron
interaction which is produced by this nonadiabatic mechanism. Thus, the matrix
elements of $H_{Cb}^0$ do not follow Eq.~\gl{eq:11} and the system represented
by $H^0$ can be described within the adiabatic approximation. The localized
states related to the atomic-like motion now are represented by adiabatic
localized functions, i.e., by the spin-dependent Wannier functions
\begin{equation}
  \label{eq:16}
  w_{{\bf T}m}({\bf r}, t) \equiv w_{im}({\bf r} - {\bf R} - \bm\rho_i,t)
\end{equation}
of the superconducting band, as defined in Eq.~(A22) of
Ref.~\cite{enhm}. The vectors ${\bf T} = {\bf R} + \bm\rho_i$ stand for
the positions of the relevant atoms, ${\bf r}$ and $t$ are the local and spin
coordinate, respectively, and $m$ denotes the quantum number of the crystal
spin in the purely electronic system.
 
Within the nonadiabatic system, the crystal spin is a conserved
quantity. Thus, also $H^0$ conserves the crystal spin of the localized
states,
\begin{equation}
  \label{eq:20}
[H^0, M(\alpha )] = 0\quad\mbox{for }\alpha \in G_M.  
\end{equation}
The purely electronic system (represented by $H^0$) is not coupled to boson
excitations that would be able to store temporarily spin-angular
momenta. Hence, $H^0$ also conserves the electron spin,
\begin{equation}
  \label{eq:21}
[H^0, S(\alpha )] = 0\quad\mbox{for }\alpha \in O(3).  
\end{equation}
Consequently, the ground state $|G^0\rangle$ of $H^0$ satisfies the equations
\begin{equation}
  \label{eq:23}
M(\alpha )|G^0\rangle = |G^0\rangle ~\mbox{ for } \alpha \in G_M
\end{equation}
and
\begin{equation}
  \label{eq:24}
S(\alpha )|G^0\rangle = |G^0\rangle ~\mbox{ for } \alpha \in O(3).
\end{equation}

In a superconducting band, a randomly chosen $N$-electron state will generally
not comply with {\em both} conditions~\gl{eq:23} {\em and}~\gl{eq:24} because
the Bloch states $|{\bf k}, m\rangle$ with crystal spin $m$ have ${\bf
k}$-dependent spin directions,
\begin{equation}
  \label{eq:30}
c_{{\bf k}qm}^{\dagger} = \sum_{s = -\frac{1}{2}}^{+\frac{1}{2}}
f_{sm}^*(q,{\bf k})c_{{\bf k}qs}^{\dagger}.
\end{equation}
The fermion operators $c_{{\bf k}qm}^{\dagger}$ and $c_{{\bf k}qs}^{\dagger}$
create Bloch electrons with crystal spin $m$ and spin $s$, respectively, and
wave vector ${\bf k}$ in the $q$th branch of the superconducting band. The
coefficients $f_{sm}(q,{\bf k})$ determine the direction of the electron spin
in the Bloch state $|{\bf k}, m\rangle$ and, consequently, form a unitary
two-dimensional matrix depending on ${\bf k}$ and $q$. In a superconducting
band, the matrix $f_{sm}(q,{\bf k})$ cannot be chosen to be independent of
${\bf k}$. Thus, only very special, if any, $N$-electron states comply with
both conditions ~\gl{eq:23} and~\gl{eq:24}.

Indeed, there exist states complying with both conditions. Let be $|G^0\rangle$
a linear combination of states
\begin{equation}
  \label{eq:8}
|Cp\rangle = \beta^{\dagger}_{{\bf k}_1 q_1}\beta^{\dagger}_{{\bf
 k}_2 q_2}\beta^{\dagger}_{{\bf k}_3 q_3}\cdots \beta^{\dagger}_{{\bf
 k}_{N/2} q_{N/2}}|0\rangle,
\end{equation}
where the new operators
\begin{equation}
  \label{eq:7}
\beta^{\dagger}_{{\bf k} q} = c_{{\bf k} q m}^{\dagger}c_{-{\bf
 k} q -m}^{\dagger} - c_{{\bf k} q -m}^{\dagger}c_{-{\bf k} q m}^{\dagger}
\end{equation}
create symmetrized Cooper pairs.

From Eq.~\gl{eq:7} it follows immediately that
\begin{equation}
  \label{eq:27}
  M(\alpha )\beta^{\dagger}_{{\bf k}q}M^{-1}(\alpha ) = \beta^{\dagger}_{{\bf
 k}q}~\mbox{ for } \alpha \in G_M, 
\end{equation}
because the operators $\beta^{\dagger}_{{\bf k}q}$ form basis functions of the
identity representation $\Gamma_1$ of $G_M$. Thus, we have
\begin{equation}
  \label{eq:26}
M(\alpha )|Cp\rangle = |Cp\rangle ~\mbox{ for } \alpha \in G_M,
\end{equation}
and, hence, Eq.~\gl{eq:23} is true. Transforming the operators
$\beta^{\dagger}_{{\bf k} q}$ in Eq.~\gl{eq:7} into the $s$ representation, we
obtain again Cooper pairs of the same form,
\begin{equation}
  \label{eq:4}
\beta^{\dagger}_{{\bf k} q} = c_{{\bf k} q s}^{\dagger}c_{-{\bf
 k} q -s}^{\dagger} - c_{{\bf k} q -s}^{\dagger}c_{-{\bf k} q s}^{\dagger},
\end{equation}
demonstrating that also Eq.~\gl{eq:24} is valid. 

Eq.~\gl{eq:4} can be deduced from Eq.~\gl{eq:7} using Eq.~\gl{eq:30} and the
time-inversion symmetry of the crystal spin,
\begin{equation}
  \label{eq:29}
Kc_{{\bf k} q m}^{\dagger}K^{-1} = v(m)c_{-{\bf k} q -m}^{\dagger},
\end{equation}
and of the electron spin,
\begin{equation}
  \label{eq:22}
 Kc_{{\bf k} q s}^{\dagger}K^{-1} = v(s)\cdot c_{-{\bf k} q -s}^{\dagger},
\end{equation}
where $K$ denotes the operator of time inversion
and
$$
v(\pm{\textstyle\frac{1}{2}}) = \pm 1.
$$

The ground state $|G^0\rangle$ of $H^0$ complies with both
conditions~\gl{eq:23} and~\gl{eq:24} only if the electrons form Cooper pairs
invariant under time-inversion. This important result demonstrates that the
nonadiabatic spin-boson interaction $H_{Cb}^{ns}$ in Eq.~\gl{eq:13} forces the
electrons in a novel way to form Cooper pairs below a certain transition
temperature. The formation of Cooper pairs is a consequence of the interplay
of two conservation laws, namely of the conservation of the electron spin and
of the conservation of the crystal spin of the localized states.

\subsection{Constraining forces required for superconducting eigenstates}
\label{sec:necessity}
To date, it is not possible to solve the Schr\"odinger equation for the
electron-boson system in a solid state. Hence, it cannot be excluded that
there exists a condition for superconducting {\em eigenstates} not yet
considered in the theory of superconductivity.

From the conditions~\gl{eq:20} and~\gl{eq:21} it follows that the interaction
term of the electronic Hamiltonian $H^0$ in Eq.~\gl{eq:15} has the form
\begin{equation}
  \label{eq:6}
  H_{Cb}^0 = \sum_{{\bf k} q, {\bf k}' q'}\langle{\bf k}, q|H_{Cb}^0|{\bf k}', q'\rangle
\beta^{\dagger}_{{\bf k} q}\beta_{{\bf k}' q'}
\end{equation}
showing that $H_{Cb}^0$ is strongly ${\bf k}$ and $s$ dependent since
\begin{equation}
  \label{eq:25}
\langle{\bf k}_{1},s_{1}, q_{1};{\bf k}_{2},s_{2}, q_{2}|H_{Cb}^0|
{\bf k}_{1}',s_{1}', q_{1}';{\bf k}_{2}',s_{2}', q_{2}'\rangle = 0
\end{equation}
for ${\bf k}_1 \neq -{\bf k}_2, {\bf k}_1' \neq -{\bf k}_2', s_1 \neq -s_2,$ or
$s_1' \neq -s_2'$.  

This Eq.~\gl{eq:25} may be interpreted as {\em condition of constraint}
indicating the existence of {\em constraining forces} in a narrow
superconducting band. Thus, Eq.~\gl{eq:25} demonstrates that $H^n$ is acting
in a special part of the Hilbert space representing a nonadiabatic system in
which constraining forces are effective in a way familiar from classical
mechanics. Below a transition temperature, these constraining forces reduce
the degrees of freedom of the electron system by forcing the electrons to form
pairs that are invariant under time inversion, i.e., by forcing the electrons
to form Cooper pairs possessing only half the degrees of freedom of unpaired
electrons.

In materials that do not possess a superconducting band, on the other hand,
constraining forces halving the degrees of freedom of the electrons do not
exist. In the classical mechanics, however, any reduction of the degrees of
freedom of any system of particles is caused by constraining forces. Hence, it
cannot be excluded that also in quantum mechanical systems any reduction of
the electronic degrees of freedom is produced by constraining forces since
quantum particles behave in some respects similar to classical particles. This
comparison of the quantum system with a classical system suggests that the
constraining forces characterized by Eq.~\gl{eq:25} {\em are required} for the
formation of Cooper pairs, i.e., they are required for the Hamiltonian to
possess superconducting eigenstates. This interpretation is corroborated by
the observation that materials not possessing a narrow, partly filled
superconducting band do not become superconducting. Thus, the author proposes
that materials which do not possess a narrow, partly filled superconducting
band, do not become superconducting even if the electrons of the considered
material are (weakly or strongly) coupled by an effective electron-electron
interaction.

It should be noted that the purely electronic system represented by the
Hamiltonian $H^0$ in Eq.~\gl{eq:15} only approximates the true nonadiabatic
system represented by the operator $H^n$ given in Eq.~\gl{eq:14}. So, the
Cooper pairs are not really {\em rigid} as suggested by the ground state
$|G^0\rangle$ of $H^0$. In the nonadiabatic system the physics of the
mechanism of constraining forces may be demonstrated more realistically in
terms of ``spring-mounted'' Cooper pairs~\cite{josi}.

\subsection{Calculation of the transition temperature}

In accordance with the generally accepted and experimentally corroborated
concept of superconductivity, the formation of Cooper pairs is mediated by
Boson excitations also within a superconducting band. Further, the constraining
forces determined by Eq.~\gl{eq:25} do not alter the energy of the electron
system but only lower the degrees of freedom of the electrons. Consequently,
the vast majority of the statements and calculations of the traditional theory
of superconductivity {\em should stay valid} in a superconducting band. In
particular, the superconducting transition temperature $T_c$ may be calculated
also in a superconducting band by a slightly modified BCS equation~\cite{bcs}
in the weak-coupling limit~\cite{josn}. A calculation of $T_c$ within the NHM
in the strong-coupling limit remains to be done. Principally, however, the
group-theoretical NHM does not provide methods to calculate the matrix
elements of $H_{Cb}^0$ in Eq.~\gl{eq:6}. Hence, it does not distinguish
between weak-coupling and strong-coupling superconductivity. It only proposes
that the equation of constraint~\gl{eq:25} is a universal condition for
superconducting eigenstates in the weak-coupling as well as in the
strong-coupling limits.
  
\setcounter{equation}{0}
\section{Superconducting band in $\text{Mg}\text{B}_2$}
\label{sec:sband2}


\begin{table}[!]
\caption{Character tables of the single-valued irreducible representations of the
  space group $P6/mmm = \Gamma_{h}D^{1}_{6h}$ (191) of MgB$_2$, as determined
  from Table 5.7 in the textbook of Bradley and Cracknell~\cite{bc}. $(j = 1,
  2, 3.)$  
\label{tab:darst2}}

\begin{tabular}[t]{cccccccccccccc}
\multicolumn{14}{c}{$\Gamma (000)$, $A (00\frac{1}{2})$}\\
 & & $E$ & $C^{\pm}_3$ & $C'_{2j}$ & $C_2$ & $C^{\pm}_6$ & $C''_{2j}$ & $I$ & $S^{\pm}_6$ & $\sigma_{dj}$ & $\sigma_h$ & $S^{\pm}_3$ & $\sigma_{vj}$\\
\hline
$\Gamma^{\pm}_1$ & $A^{\pm}_1$ & 1 & 1 & 1 & 1 & 1 & 1 & $\pm 1$ & $\pm 1$ & $\pm 1$ &
$\pm 1$ & $\pm 1$ & $\pm 1$\\
$\Gamma^{\pm}_2$ & $A^{\pm}_2$ & 1 & 1 & -1 & 1 & 1 & -1 & $\pm 1$ & $\pm 1$ & $\mp 1$ & $\pm 1$ & $\pm 1$ & $\mp 1$\\
$\Gamma^{\pm}_3$ & $A^{\pm}_3$ & 1 & 1 & 1 & -1 & -1 & -1 & $\pm 1$ & $\pm 1$ & $\pm 1$ & $\mp 1$ & $\mp 1$ & $\mp 1$\\
$\Gamma^{\pm}_4$ & $A^{\pm}_4$ & 1 & 1 & -1 & -1 & -1 & 1 & $\pm 1$ & $\pm 1$ & $\mp
1$ & $\mp 1$ & $\mp 1$ & $\pm 1$\\
$\Gamma^{\pm}_5$ & $A^{\pm}_5$ & 2 & -1 & 0 & -2 & 1 & 0 & $\pm 2$ & $\mp 1$ & 0 & $\mp 2$ &
$\pm 1$ & 0\\
$\Gamma^{\pm}_6$ & $A^{\pm}_6$ & 2 & -1 & 0 & 2 & -1 & 0 & $\pm 2$ & $\mp 1$ & 0 & $\pm 2$ & $\mp 1$ & 0\\
\hline\\
\end{tabular}\hspace{1cm}
\begin{tabular}[t]{cccccccc}
\multicolumn{8}{c}{$H (\overline{\frac{1}{3}}\frac{2}{3}\frac{1}{2})$, $K
  (\overline{\frac{1}{3}}\frac{2}{3}0)$}\\
 & & $E$ & $\sigma_h$ & $S^{\pm}_3$ & $C^{\pm}_3$ & $C''_{2j}$ & $\sigma_{dj}$\\
\hline
$H_1$ & $K_1$ & 1 & 1 & 1 & 1 & 1 & 1\\
$H_2$ & $K_2$ & 1 & 1 & 1 & 1 & -1 & -1\\
$H_3$ & $K_3$ & 1 & -1 & -1 & 1 & 1 & -1\\
$H_4$ & $K_4$ & 1 & -1 & -1 & 1 & -1 & 1\\
$H_5$ & $K_5$ & 2 & 2 & -1 & -1 & 0 & 0\\
$H_6$ & $K_6$ & 2 & -2 & 1 & -1 & 0 & 0\\
\hline\\
\end{tabular}\hspace{1cm}
\begin{tabular}[t]{cccccccccc}
\multicolumn{10}{c}{$M (0\frac{1}{2}0)$, $L (0\frac{1}{2}\frac{1}{2})$}\\
 & & $E$ & $C_2$ & $C''_{21}$ & $C'_{21}$ & $I$ & $\sigma_h$ & $\sigma_{v1}$ &
 $\sigma_{d1}$\\ 
\hline
$M^{\pm}_1$ & $L^{\pm}_1$ & 1 & 1 & 1 & 1 & $\pm 1$ & $\pm 1$ & $\pm 1$ & $\pm 1$\\
$M^{\pm}_2$ & $L^{\pm}_2$ & 1 & -1 & 1 & -1 & $\pm 1$ & $\mp 1$ & $\pm 1$ & $\mp 1$\\
$M^{\pm}_3$ & $L^{\pm}_3$ & 1 & 1 & -1 & -1 & $\pm 1$ & $\pm 1$ & $\mp 1$ & $\mp 1$\\
$M^{\pm}_4$ & $L^{\pm}_4$ & 1 & -1 & -1 & 1 & $\pm 1$ & $\mp 1$ & $\mp 1$ & $\pm 1$\\
\hline\\
\end{tabular}
\end{table}


We show now that the energy band denoted in Fig.~\ref{fig:bandstr2}
by the bold line is a superconducting band. It is labeled by the representations
\begin{equation}
\label{eq:1b}
\Gamma^-_5,\Gamma^+_6;\ K_6;\ M^-_3, M^+_2;\ A^+_5, A^-_6;\ L^+_1, L^-_4;\ H_5. 
\end{equation}

\begin{table}[!]
\caption{Character tables of the double-valued irreducible representations of the
  space group $\Gamma_{h}D^{1}_{6h}$ of MgB$_2$, as determined
  from Table 6.13 of Bradley and Cracknell~\protect\cite{bc}. $(j = 1, 2, 3.)$   
\label{tab:zdarst2}}
\begin{tabular}[t]{ccccccccccc}
\multicolumn{11}{c}{$\Gamma (000)$,\ $A (00\frac{1}{2})$}\\
 &  & $$ & $$ & $$ & $$ & $$ & $$ & $\overline{C}_2$ & $\overline{C}'_{2j}$ & $\overline{C}''_{2j}$\\
 &  & $E$ & $\overline{E}$ & $C^{\pm}_6$ & $\overline{C}^{\pm}_6$ &
 $C^{\pm}_3$ & $\overline{C}^{\pm}_3$ & $C_2$ & $C'_{2j}$ & $C''_{2j}$\\
\hline
$\Gamma^{\pm}_7$ & $A^{\pm}_7$ & 2 & -2 & 0 & 0 & -2 & 2 & 0 & 0 & 0\\
$\Gamma^{\pm}_8$ & $A^{\pm}_8$ & 2 & -2 & $\sqrt{3}$ & $-\sqrt{3}$ & 1 & -1 &
0 & 0 & 0\\
$\Gamma^{\pm}_9$ & $A^{\pm}_9$ & 2 & -2 & $-\sqrt{3}$ & $\sqrt{3}$ & 1 & -1 &
0 & 0 & 0\\
\hline\\
\end{tabular}\hspace{.5cm}
\begin{tabular}[t]{ccccccccccc}
\multicolumn{11}{c}{$\Gamma (000)$,\ $A (00\frac{1}{2})$\quad \emph{(continued)}}\\
 &  & $$ & $$ & $$ & $$ & $$ & $$ & $\overline{\sigma}_h$ & $\overline{\sigma}_{dj}$ & $\overline{\sigma}_{vj}$\\
 &  & $I$ & $\overline{I}$ & $S^{\pm}_3$ & $\overline{S}^{\pm}_3$ & $S^{\pm}_6$ & $\overline{S}^{\pm}_6$ & $\sigma_h$ & $\sigma_{dj}$ & $\sigma_{vj}$\\
\hline
$\Gamma^{\pm}_7$ & $A^{\pm}_7$ &  $\pm 2$ & $\mp 2$ & 0 & 0 & $\mp 2$ & $\pm 2$ & 0 & 0 & 0\\ 
$\Gamma^{\pm}_8$ & $A^{\pm}_8$ & $\pm 2$ & $\mp 2$ & $\pm\sqrt{3}$ & $\mp\sqrt{3}$ & $\pm 1$ & $\mp 1$ & 0 & 0 & 0\\
$\Gamma^{\pm}_9$ & $A^{\pm}_9$ & $\pm 2$ & $\mp 2$ & $\mp\sqrt{3}$ & $\pm\sqrt{3}$ & $\pm 1$ & $\mp 1$ & 0 & 0 & 0\\
\hline\\
\end{tabular}
\begin{tabular}[t]{ccccccccccc}
\multicolumn{11}{c}{$H (\overline{\frac{1}{3}}\frac{2}{3}\frac{1}{2})$,\ $K
  (\overline{\frac{1}{3}}\frac{2}{3}0)$}\\ 
 &  & $$ & $$ & $$ & $$ & $$ & $$ & $\overline{\sigma}_h$ & $\overline{C}''_{2j}$ & $\overline{\sigma}_{dj}$\\
 &  & $E$ & $\overline{E}$ & $S^{\pm}_3$ & $\overline{S}^{\pm}_3$ & $C^{\pm}_3$ & $\overline{C}^{\pm}_3$ & $\sigma_h$ & $C''_{2j}$ & $\sigma_{dj}$\\
\hline
$H_7$ & $K_7$ & 2 & -2 & 0 & 0 & -2 & 2 & 0 & 0 & 0\\
$H_8$ & $K_8$  & 2 & -2 & $\sqrt{3}$ & $-\sqrt{3}$ & 1 & -1 & 0 & 0 & 0\\
$H_9$ & $K_9$  & 2 & -2 & $-\sqrt{3}$ & $\sqrt{3}$ & 1 & -1 & 0 & 0 & 0\\
\hline\\
\end{tabular}\hspace{.5cm}
\begin{tabular}[t]{cccccccccccc}
\multicolumn{12}{c}{$M (0\frac{1}{2}0)$,\ $L (0\frac{1}{2}\frac{1}{2})$}\\
 &  & $$ & $$ & $\overline{C}_2$ & $\overline{C}''_{21}$ & $\overline{C}'_{21}$ & $$ & $$ & $\overline{\sigma}_h$ & $\overline{\sigma}_{v1}$ & $\overline{\sigma}_{d1}$\\
 &  & $E$ & $\overline{E}$ & $C_2$ & $C''_{21}$ & $C'_{21}$ & $I$ & $\overline{I}$ & $\sigma_h$ & $\sigma_{v1}$ & $\sigma_{d1}$\\
\hline
$M^{\pm}_5$ & $L^{\pm}_5$ & 2 & -2 & 0 & 0 & 0 & $\pm$2 & $\mp$2 & 0 & 0 & 0\\
\hline\\
\end{tabular}

\end{table}


Table~\ref{tab:atomwf2} (a) lists all the four bands in $\text{Mg}\text{B}_2$
whose Bloch functions can be unitarily transformed into symmetry-adapted and
best localized Wannier functions situated on the B atoms. Each band consists
of two branches because there are two B atoms in the unit cell. While the
representations~\gl{eq:1b} coincide with the representations of band 4 in
Table~\ref{tab:atomwf2} (a) at points $L, M, K,$ and $H$, the representations
at points $\Gamma$ and $A$ are different. Hence, we {\em cannot} represent the
Bloch functions of this band by symmetry-adapted and best localized
(spin-\emph{in}dependent) Wannier functions centered on the B sites.


\begin{table}[!]
  \caption{
    Compatibility relations between the single-valued (upper row) and
    double-valued (lower row) representations of the space group
    $\Gamma_{h}D^{1}_{6h}$ of MgB$_2$. 
\label{tab:comprel}
}
\begin{tabular}[t]{cccccc}
\multicolumn{6}{c}{$\Gamma$, $A$}\\
\hline
$R^{\pm}_1$ & $R^{\pm}_2$ & $R^{\pm}_3$ & $R^{\pm}_4$ & $R^{\pm}_5$ & $R^{\pm}_6$ \\
$R^{\pm}_8$ & $R^{\pm}_8$ & $R^{\pm}_9$ & $R^{\pm}_9$ & $R^{\pm}_7$ +
$R^{\pm}_8$ & $R^{\pm}_7$ + $R^{\pm}_9$ \\
\hline\\
\end{tabular}\hspace{1cm}
\begin{tabular}[t]{cccccc}
\multicolumn{6}{c}{$H$, $K$}\\
\hline
$R_1$ & $R_2$ & $R_3$ & $R_4$ & $R_5$ & $R_6$\\
$R_8$ & $R_8$ & $R_9$ & $R_9$ & $R_7$ + $R_9$ & $R_7$ + $R_8$\\
\hline\\
\end{tabular}\hspace{1cm}
\begin{tabular}[t]{c}
\multicolumn{1}{c}{$L$, $M$}\\
\hline
$R^{\pm}_i$, $i = 1,2,3,4$\\
$R^{\pm}_5$\\
\hline\\
\end{tabular}
\footnotetext{
Notes to Table~\ref{tab:comprel}
\begin{enumerate}
\item The single-valued and double-valued representations are listed in
  Tables~\ref{tab:darst2} and~\ref{tab:zdarst2}, respectively. 
\item Each column lists the double-valued representation $R_i\times D_{1/2}$
    below the single-valued representation $R_i$.
\end{enumerate}
}
\end{table}


The situation is changed when we replace the single-valued representations
$R^{\pm}_i$ by the corresponding double-valued representations
$R^{\pm}_i\times D_{1/2}$ listed in Table~\ref{tab:comprel}. The energy band
with the representations~\gl{eq:1b} now is characterized by the double-valued
representations
\begin{equation}
\begin{array}{l}
  \label{eq:2}
  \Gamma^-_7,\underline{\Gamma^-_8},\Gamma^+_7,\underline{\Gamma^+_9};\ 
  \underline{K_7} , \underline{K_8};\ \underline{M^-_5}, \underline{M^+_5};\\
  A^+_7, \underline{A^+_8}, A^-_7,\underline{A^-_9};\ \underline{L^+_5},
  \underline{L^-_5};\ \underline{H_7}, \underline{H_9}.
\end{array}   
\end{equation}
The underlined representations form band 3 in Table~\ref{tab:atomwf2}
(b). Hence, the associated Bloch functions can be unitarily transformed into
symmetry-adapted and best localized {\em spin-dependent} Wannier functions
situated on the B sites.


\begin{table}[!]
\caption{
Single- and double-valued representations of all the energy bands in
MgB$_2$ with symmetry-adapted and optimally 
localized (spin-dependent) Wannier functions centered at the B atoms. 
\label{tab:atomwf2}}
\begin{tabular}[t]{ccccccc}
\multicolumn{7}{c}{\em (a)\quad Single-valued representations}\\
& $\Gamma$ & $L$ & $M$ & $A$ & $K$ & $H$\\
\hline
Band 1\ \ & $\Gamma^+_1$ + $\Gamma^-_3$ & $L^+_2$ + $L^-_3$ & 
$M^+_1$ + $M^-_4$ & $A^+_4$ + $A^-_2$ & $K_5$ & $H_6$\\
Band 2\ \ & $\Gamma^+_2$ + $\Gamma^-_4$ & $L^+_4$ + $L^-_1$ & $M^+_3$
 + $M^-_2$ & $A^+_3$ + $A^-_1$ & $K_5$ & $H_6$\\
Band 3\ \ & $\Gamma^+_3$ + $\Gamma^-_1$ & $L^+_3$ + $L^-_2$ & $M^+_4$
 + $M^-_1$ & $A^+_2$ + $A^-_4$ & $K_6$ & $H_5$\\
Band 4\ \ & $\Gamma^+_4$ + $\Gamma^-_2$ & $L^+_1$ + $L^-_4$ & $M^+_2$
 + $M^-_3$ & $A^+_1$ + $A^-_3$ & $K_6$ & $H_5$\\
\hline\\
\end{tabular}
\begin{tabular}[t]{cccc}
\multicolumn{4}{c}{\em (b)\quad Double-valued representations}\\
& $\Gamma$ & $M$ & $A$\\
\hline
Band 1\ \ & $\Gamma^+_7$ + $\Gamma^-_7$ & $M^+_5$ + $M^-_5$ & 
$A^+_7$ + $A^-_7$\\
Band 2\ \ & $\Gamma^+_8$ + $\Gamma^-_9$ & $M^+_5$ + $M^-_5$ & $A^+_9$
 + $A^-_8$\\
Band 3\ \ & $\Gamma^+_9$ + $\Gamma^-_8$ & $M^+_5$ + $M^-_5$ & $A^+_8$
 + $A^-_9$\\
\hline\\
\end{tabular}\hspace{.5cm}
\begin{tabular}[t]{cccc}
\multicolumn{4}{c}{\em (b)\quad (continued)}\\
& $L$ & $K$ & $H$\\
\hline
Band 1\ \ & $L^+_5$ + $L^-_5$ & $K_8$ + $K_9$ & $H_8$ + $H_9$\\
Band 2\ \ & $L^+_5$ + $L^-_5$ & $K_7$ + $K_9$ & $H_7$ + $H_8$\\
Band 3\ \ & $L^+_5$ + $L^-_5$ & $K_7$ + $K_8$ & $H_7$ + $H_9$\\
\hline\\
\end{tabular}
\footnotetext{
Notes to Table~\ref{tab:atomwf2}
\begin{enumerate}
\item The bands 2 and 3 in Table~\ref{tab:atomwf2} (b) form
  superconducting bands.
\item Band 1 of Table~\ref{tab:atomwf2} (b) is \emph{not} a superconducting
  band because the representations $\Gamma^{\pm}_7$ cannot be written in the
  form $R\times D_{1/2}$ where $R$ stands for any one-dimensional
  single-valued representation and $D_{1/2}$ denotes the two-dimensional
  double-valued representation of the three-dimensional rotation group
  $O(3)$. The form $R\times D_{1/2}$ of the representations $\Gamma^{\pm}_8$
  and $\Gamma^{\pm}_9$ belonging to bands 2 and 3 in Table~\ref{tab:atomwf2}
  (b) ensures that the spin-dependent Wannier functions transform under the
  space group operations like spin functions, see Eq.~A(28) of
  Ref.~\cite{enhm}. 
\item The single-valued and double-valued representations are listed in
  Tables~\ref{tab:darst2} and~\ref{tab:zdarst2}, respectively. 
\item Each row defines one band consisting of two branches, because there
  are two B atoms in the unit cell.
\item The bands are determined by Eq.~(23) of Ref.~\protect\cite{josla2cuo4}. 
\item Assume a band of the symmetry in any row of this table to exist in the
  band structure of MgB$_2$.
  Then the Bloch functions of this band can be unitarily transformed into
  Wannier functions that are
\begin{itemize}
\item localized as well as possible; 
\item centered at the B atoms; and
\item symmetry-adapted to the space group $\Gamma_{h}D^{1}_{6h}$ of MgB$_2$. 
\end{itemize}
These Wannier function are usual (spin-independent) Wannier function if the
considered band is characterized by the single-valued representations (a). They
are spin dependent if the band is characterized by the double-valued
representations (b).
\end{enumerate}
}
\end{table}


These Wannier functions are only weakly spin dependent since
they do not strongly differ from the Wannier function belonging to band 4 in
Table~\ref{tab:atomwf2} (a). It is only the representations $\Gamma^+_4$ +
$\Gamma^-_2$ and $A^+_1$ + $A^-_3$ of band 4 in Table~\ref{tab:atomwf2} (a)
which do not belong to the representations~\gl{eq:1b}.

Between $A$ and $L$ the superconducting band jumps from the upper to the lower
band. This small jump is allowed within the NHM because the Bloch
functions of both bands belong to the same double-valued representation and
the jump does not cross the Fermi level.

\setcounter{equation}{0}
\section{Discussion}
\label{sec:disc}
This paper shows that the intermetallic superconducting compound MgB$_2$
possesses a narrow, roughly half-filled ``superconducting band'' in its band
structure, see Fig.~\ref{fig:bandstr2}. In addition to the previous
observations about elemental superconductors, non-superconductors,
YBa$_2$Cu$_3$O$_7$, and La$_2$CuO$_4$ (as mentioned in Sec.~\ref{sec:intro})
this result provides further evidence that superconducting bands are a general
feature of both weak-coupling and strong-coupling superconductors. Hence, the
author proposes that in any material Cooper pairs are stabilized by the
constraining forces determined by the condition of constraint~\gl{eq:25} and
generated by the nonadiabatic condensation mechanism characterized by
Eq.~\gl{eq:11}. These constraining forces are effective only in narrow
superconducting bands and are proposed to be required that the Hamiltonian of
the electron-boson system possesses superconducting \emph{eigenstates}.

The symmetry of the Bloch functions of the \emph{entire} superconducting band
determines the symmetry and spin dependence of the nonadiabatic localized
states. The nonadiabatic condensation mechanism characterized by
Eq.~\gl{eq:11}, on the other hand, is produced by the Coulomb
\emph{correlation} energy of the electrons \emph{near the Fermi level}, in
accordance with the generally accepted concept that correlated conduction
electrons are responsible for superconductivity.

Further, in accordance with the general belief, also in superconducting bands
the formation of Cooper pairs is mediated by boson excitations. In
superconducting bands, however, the pair formation is mediated by the {\em
  energetically lowest} boson excitations of the crystal that possess the
crystal-spin angular momentum $1\cdot\hbar$ and are sufficiently stable to
transport it through the crystal.  These ``crystal-spin-1'' bosons are
localized excitations $|{\bf T}, l\rangle$ (with $l = -1, 0, +1$ labeling the
three directions of the crystal spin and ${\bf T}$ denoting a lattice point) of
well-defined symmetry~\cite{es,ehtc} which propagate as Bloch waves (with the
crystal momentum $\hbar\cdot{\bf k}$) through the crystal.

The $|{\bf T}, l\rangle$ are generated during spin-flip processes in the
superconducting band and must carry off the surplus crystal-spin
angular-momenta generated at these processes. This spin-boson mechanism
suggests that the $|{\bf T}, l\rangle$ are coupled phonon-plasmon modes: In a
first step the atomic-like electrons in the superconducting band transmit
their angular momenta to the core electrons by generating a plasmon-like
vibration of the core electrons against the atoms. In a second step, these
plasmon-like excitations generate phonon-like vibrations of lower energy if
crystal-spin-1 phonons are sufficiently stable in the considered material.

Thus, the author proposes that the $|{\bf T}, l\rangle$ are coupled
phonon-plasmon modes which have dominant phonon character in the isotropic
lattices of the transition elements and, hence, confirm the electron-phonon
mechanism that enters the BCS theory~\cite{bcs} in these materials~\cite{es}. 
However, phonon-like excitations are not able to transport crystal-spin
angular-momenta within the two-dimensional copper-oxygen layers of
the cuprates, see Ref.~\cite{ehtc} for preliminary ideas to this
problem.  Within two-dimensional layers, the $|{\bf T}, l\rangle$
necessarily are energetically higher lying excitations of dominant plasmon
character.  This clear dependence of stable crystal-spin-1 bosons on the
properties of the lattice suggests that they are (at least partially)
responsible for the special properties of the layered superconductors, i.e.,
their strong-coupling features and their high transition temperatures.

Also MgB$_2$ contains two-dimensional hexagonal nets of B atoms. Thus, also in
this layered material crystal-spin-1 \emph{phonons} will be less stable than
in the isotropic lattices of the transition elements. The balance between the
phonon and plasmon character of stable crystal-spin-1 excitations is shifted
towards the plasmon character leading to a higher transition temperature and
to the experimentally established~\cite{hinks,budko,brotto} reduced isotope
effect.

The superconducting band in MgB$_2$ is composed of $\sigma$- and $\pi$-bands
in accordance with the two-band model of
superconductivity~\cite{koshelev,mazin_1} in this material characterized by
$\sigma$- and $\pi$-bands associated with different parts of the Fermi
surface~\cite{mazin_2}. The part of the Fermi surface of the superconducting
band enclosing the points $\Gamma$ and $A$ has $\sigma$ character, the other
parts have $\pi$ character.

In the band structures of the two-dimensional superconductors
YBa$_2$Cu$_3$O$_7$~\cite{ybacuo7}, MgB$_2$ (this paper), and
La$_2$CuO$_4$~\cite{josla2cuo4} I found a new feature of the superconducting
bands: the related spin-dependent Wannier functions are only {\em weakly} spin
dependent. I believe that this weak spin dependence is an additional condition
for stable two-dimensional high-$T_c$ superconducting states. This question
requires further theoretical consideration and further examination of the band
structures of high-$T_c$ superconductors.

\acknowledgements{%
  I am indebted to Ove Jepsen for providing me with the band-structure data of
  MgB$_2$. I thank Horst Strunk for his support of my work and Ernst Helmut
  Brandt for critical and helpful comments on the manuscript.}


\begin{thebibliography}{26}
\bibitem[{bem()}]{bem2}
\bibinfo{note}{With free atoms, the nonadiabatic motion of the nuclei occurs in
  such a way that the total centre of mass stays at rest. In a solid state,
  consisting of bound atoms, this restriction does not hold.}
\bibitem[{\citenamefont{Kr{\"u}ger}(2001{\natexlab{c}})}]{josn}
\bibinfo{author}{\bibfnamefont{E.}~\bibnamefont{Kr{\"u}ger}},
  \bibinfo{journal}{J. Supercond.} \textbf{\bibinfo{volume}{14(4)}},
  \bibinfo{pages}{469} (\bibinfo{year}{2001}{\natexlab{c}}), \bibinfo{note}{it
  should be noted that in this paper the term ``superconducting band'' was
  abbreviated by ``$\sigma$-band''}.
\end{thebibliography}
\end{document}